\RequirePackage{amsmath}
\documentclass[12pt,a4paper,final]{iopart}

\usepackage{graphicx}

\usepackage[makeroom]{cancel}
\usepackage{dsfont}

\usepackage{fancyhdr}
 
\usepackage{hyperref}
\hypersetup{
  colorlinks   = true, 
  urlcolor     = blue, 
  linkcolor    = blue,
  citecolor   = blue 
}

\newcommand{\bra}[1]{\langle #1|}
\newcommand{\ket}[1]{|#1\rangle}

\begin{document}

\title{Carrier-free Raman manipulation of trapped neutral atoms}

\author{Ren\'{e} Reimann, Wolfgang Alt, Tobias Macha, Dieter Meschede, Natalie Thau, Seokchan Yoon, and Lothar Ratschbacher}
\address{Institut f\"ur Angewandte Physik der Universit\"at Bonn, Wegelerstrasse 8, 53115 Bonn, Germany}

\eads{\mailto{ratschbacher@iap.uni-bonn.de}}

\begin{abstract}
We experimentally realize an enhanced Raman control scheme for neutral atoms that features an intrinsic suppression of the two-photon carrier transition, but retains the sidebands which couple to the external degrees of freedom of the trapped atoms. This is achieved by trapping the atom at the node of a blue detuned standing wave dipole trap, that acts as one field for the two-photon Raman coupling. The improved ratio between cooling and heating processes in this configuration enables a five times lower fundamental temperature limit for resolved sideband cooling. We apply this method to perform Raman cooling to the two-dimensional vibrational ground state and to coherently manipulate the atomic motion. The presented scheme requires minimal additional resources and can be applied to experiments with challenging optical access, as we demonstrate by our implementation for atoms strongly coupled to an optical cavity. 
\end{abstract}

\vspace{2pc}
\section{Introduction}

Trapped single atoms and atomic ensembles represent a versatile platform for the investigation and application of quantum physics with an extraordinary level of control. The manipulation of the quantum states of localized neutral atoms has in recent years formed the basis for fundamental studies of quantum mechanics~\cite{purdy2010, brahms2012}, high precision metrology~\cite{derevianko2011} and the implementation of quantum information~\cite{kimble2008, ritter2012} and quantum simulation protocols~\cite{block2012}. 

Crucial to many of these and future experiments is the capability to efficiently control the motional degree of freedom of the atoms. In order to localize and prepare neutral atoms with high probability in their motional ground states two different approaches exist. Evaporative cooling of large atomic ensembles has been the established route towards ultracold temperatures~\cite{anderson1995} and remains essential for achieving the most dense atomic ensembles with lowest entropy~\cite{greiner2002}. The need for collisional thermalization and the inherent particle loss, however, result in long preparation times and can limit the measurement duty cycle. 

In experiments with a smaller number of atoms Raman sideband cooling has emerged as an alternative for preparing quantum motional states of trapped neutral atoms. Using this method strongly confined neutral atoms can directly be laser cooled into the vibrational ground state of their respective conservative trapping potentials, as has recently been shown with single neutral atoms in optical tweezers~\cite{kaufman2012,thompson2013} and cavities~\cite{boozer2006, reiserer2013}. These advances are opening exciting prospects for quantum information science with trapped neutral atoms, which benefit from the convenient scaling properties of optically generated potentials~\cite{schrader2004, lengwenus2010, piotrowicz2013, nogrette2014}. The lossless recooling on millisecond timescales adds to the growing capabilities of neutral atoms, which include the deterministic preparation of single-atom~\cite{grunzweig2010} and two-atom particle number Fock states~\cite{ebert2014} at single trap sites, the demonstration of two-atom Rydberg gates~\cite{gaetan2009, urban2009}, integration with optical resonators~\cite{boca2004, tieke2014}, and fast lossless spin detection~\cite{fuhrmanek2011,gibbons2011}. 

For neutral atoms the conditions for robust Raman cooling, i.e. the presence of resolved motional sidebands of appropriate coupling strength, can be challenging to fulfill. This is particularly true for setups with restricted optical access and unconventional optical potentials, such as cavity QED systems~\cite{boozer2006,boozer2008} and micro-array traps~\cite{nogrette2014}, where tight harmonic confinement and the integration of additional Raman beams can be harder to achieve. Here, we demonstrate a scheme for the Raman manipulation of trapped neutral atoms that exhibits a strong suppression of the carrier (recoil-free) transition in the driven two-photon transfer and show how it can benefit the Raman cooling limit. Furthermore, we apply the technique to perform two-dimensional ground state cooling of atoms strongly coupled to an optical high-finesse cavity and to investigate their heating rate. Our results fundamentally extend the regime of motional coupling between light and atoms that is accessible in experiments. Previously the ratio of motional sideband coupling to carrier coupling strength has been varied by changing the propagation direction of the two Raman beams. Pure carrier coupling is obtained for co-propagating beams, whereas maximum motional coupling is obtained for counter-propagating beams. In our experiment we go beyond this modification of the effective Lamb-Dicke factor and achieve exclusive coupling to motional sidebands through a complete suppression of carrier coupling.   

\begin{figure*}[ht]
\centering
    \includegraphics[width=1\columnwidth]{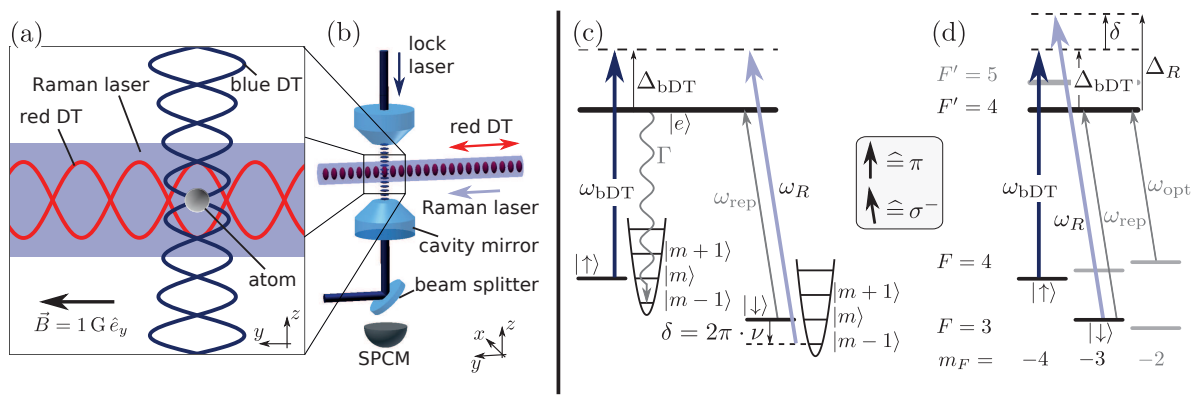}
	\caption[Level Scheme]{\label{fig:figNew}(Color) Experimental setup. (a) A single cesium atom is trapped in the optical lattice potential of a blue detuned and a red detuned standing wave dipole trap (DT). 
(b) The atom is positioned inside an optical high-finesse Fabry-P\'{e}rot cavity. The resonantly enhanced cavity lock laser field acts as the blue detuned standing wave trap for the atoms. (c) Illustration of a Raman cooling cycle driven by the blue detuned dipole trap ($\omega_\textrm{bDT}$) and the Raman laser ($\omega_R$). (d) Details of the levels and transitions in $^{133}$Cs involved in Raman manipulation and cooling. We define the two-photon detuning $\delta$ as the difference between the one-photon detunings, $\delta = \Delta_R - \Delta_\textrm{bDT}$.} 
\end{figure*}

\section{Methods}

In our experiment a single neutral cesium atom is captured from background gas by a high-gradient magneto-optical trap (MOT) and loaded into a red detuned standing wave trap at $\lambda_{\textrm{rDT}}=1030\,$nm. By using the red dipole trap as a conveyor belt the atom is translated into an orthogonal standing wave that is formed by the blue detuned ($\lambda_{\textrm{bDT}}=845.5\,$nm) locking light of an optical high-finesse Fabry-P\'{e}rot cavity (for details  see~\cite{khudaverdyan2008}). In the combined optical potential the atom has oscillation frequencies $\nu_y \approx 400\:$kHz, $\nu_z \approx 200\:$kHz and $\nu_x \approx 2\:$kHz along the red dipole trap (conveyor belt) axis, the blue dipole trap (cavity) axis and the orthogonal trap axis, respectively. The geometry of the optical potential, the orientation of the applied magnetic bias field of $1.0\,$Gauss and the position of additional laser fields in the experiment are sketched in Fig.~\ref*{fig:figNew}. 

In order to achieve coherent two-photon coupling between the $F=3$ and $F=4$ manifold (in the following we use ${\ket{\!\downarrow}}= \ket{F=3,m_F=-3}$, ${\ket{\!\uparrow}}= \ket{F=4,m_F=-4}$) of the $6\,^2\textrm{S}_{1/2}$ electronic ground state, we address the atom by a single Raman laser beam that propagates along the $y$-axis. This weak Raman light is phase locked to the blue detuned strong dipole trap light ($845.5\,$nm) with a tunable frequency offset around the hyperfine splitting $\omega_R-\omega_\textrm{bDT} \approx 2\pi \cdot 9.2\:$GHz. The second field in the two-photon Raman coupling in our experiment is hence provided by the always present blue dipole trap light~\cite{boozer2008}, in contrast to the more common implementation via a second Raman beam~\cite{kaufman2012,thompson2013}. The choice of blue detuned dipole light, where atoms are confined around the intensity zeros of the trapping light, allows us to operate the optical trap at a detuning of $\Delta_\textrm{bDT} \approx 2\pi \cdot 3$\:THz from the atomic resonance with minimal light shifts of atomic levels and low spontaneous photon scattering rates~\cite{Note1}. More importantly, however, the blue detuning provides the mechanism for the intrinsic suppression of the Raman carrier coupling investigated in this report.

\section{Results and Discussion}

Intuitively the effect can be understood by considering the local Rabi frequency for a two-photon transition, $ (\Omega_{\textrm{bDT}} \Omega_{R})/\Delta$, 
where $\Omega_\textrm{bDT}$, $\Omega_R$, and $\Delta \approx  \Delta_\textrm{bDT} \approx \Delta_R$ denote the local single-photon Rabi frequencies and detunings from resonance (cf. Fig.~\ref*{fig:figNew}d). 
For atoms fixed at the intensity zero of the blue detuned trap light the single-photon Rabi frequency $\Omega_\textrm{bDT}$ and hence the local two-photon Rabi frequency  will vanish. 

If we formally take into account the external degree of freedom of the trapped atoms, the resonant couplings between the spin-motional states are described by 
\begin{equation}
\Omega_{\uparrow m_y m_z,\downarrow m_y' m_z'} = \Omega_0 |\bra{\uparrow \!m_y' m_z'}\,\sin(k_{z}\hat{z})\,e^{ik_{y}\hat{y}}\,\hat{\sigma}^{\dagger}\,{\ket{\!\downarrow \! m_y m_z}}|  
\label{eq:eq1}
\end{equation}
where $m_y,m_z,m_y',m_z'$ denote the motional quantum numbers of the initial and the final state, $\hat{y}, \hat{z}$ are the position operators along the $y$- and $z$-axis and $k_{y}=2\pi/\lambda_{\textrm{rDT}}, k_{z}=2\pi/\lambda_\textrm{bDT}$ are the wave vectors of the red and blue detuned dipole trap fields. $\hat{\sigma}^{\dagger} = {\ket{\!\uparrow}}{\bra{\downarrow\!}}$ represents the spin raising operator, and the bare two-photon Rabi frequency with an approximate experimental value of $\Omega_0\approx 2\pi \cdot 0.3\,$MHz summarizes the dependency on laser powers, detuning and internal states.

In the Lamb-Dicke regime we can approximate the expression describing the geometry of the light fields in Eqn.~\ref{eq:eq1} and rewrite it in terms of harmonic oscillator raising $\hat{b}_z^\dagger,\hat{b}_y^{\dagger}$ and lowering operators $\hat{b}_z,\hat{b}_y$ as 
\begin{align}
\label{eq:eq2}
\sin(k_{z}\hat{z})\,e^{ik_{y}\hat{y}} &\approx (k_z \hat{z})\,(\hat{\mathds{1}}_y+i\,k_y \hat{y})  \\
&=\eta_z (\hat{b}_z^{\dagger}+\hat{b}_z)+ i\,\eta_y \eta_z (\hat{b}_z^{\dagger}\hat{b}_y^{\dagger} +\hat{b}_z \hat{b}_y^{\dagger}+\hat{b}_z^{\dagger}\hat{b}_y+\hat{b}_z \hat{b}_y).\notag 
\end{align}
Here, $\eta_y$ and $\eta_z$ denote the Lamb-Dicke factors along the $y$- and $z$-direction with experimental values of about $0.1$. Eqn.~\ref{eq:eq2} prescribes the selection rule $\Delta m_{z}=\pm 1$ for the standing wave Raman field configuration \cite{Note2}, i.e. carrier transitions with $\Delta m_{z}=0$ are suppressed. The first order sidebands of the motion along the $z$-axis scale with $\Omega_0 \eta_z$ and the sidebands coupling the motion along the $y$- and $z$-axis scale with $\Omega_0 \eta_y \eta_z$. 
 
We start our experimental investigation by mapping out the two-photon spectrum of atoms trapped in the crossed standing wave potentials to localize and identify  accessible Raman transitions. Following a successful single atom loading event, the measurement sequence initializes the atoms with high fidelity in the state ${\ket{\!\!\uparrow}}$ by a $5\,$ms long optical pumping pulse of repump ($\omega_{\textrm{rep}}$) and optical pump ($\omega_{\textrm{opt}}$) light. These two beams are $\sigma^-$-polarized and resonant with the $F=3\rightarrow F'=4$ and the $F=4\rightarrow F'=4$ of the D$_2$-line, respectively (see Fig.~\ref*{fig:figNew}d). A single Raman laser pulse of $1\,$ms duration and about $1\,$mW power (waist radius $60\,\mu$m) is applied to the atoms and followed by cavity-assisted readout of the hyperfine state of the atom~\cite{bochmann2010,boozer2006}. The fast and non-destructive state detection also provides cavity-cooling and allows us to repeat the $70\,$ms long interrogation sequence up to $100$ times with the same atom, after which a new atom is loaded (see Appendix \hyperref[AppendixA]{A}). 
\begin{figure} 
\centering
\includegraphics[width=0.5\columnwidth]{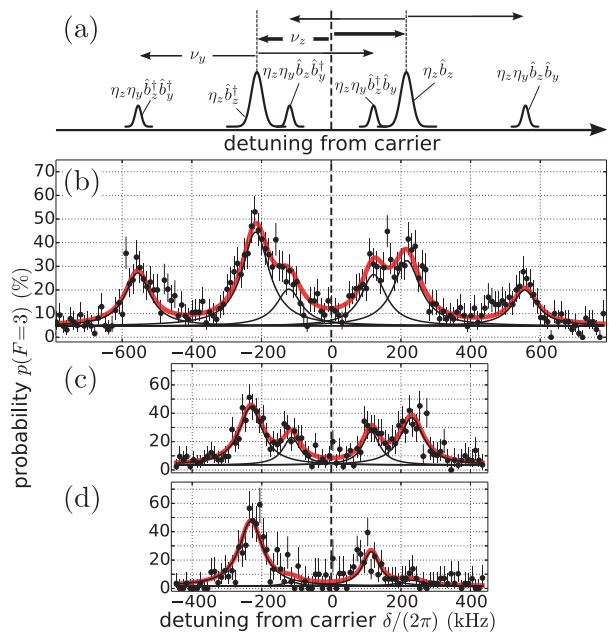}
\caption[Level Scheme]{\label{fig:fig2}(Color) Raman spectra with carrier suppression. (a) Illustration of the transitions allowed by the Raman field geometry (cf. Eqn.~\ref{eq:eq2}). (b) Complete Raman spectrum after cavity cooling. Error bars indicate one standard deviation uncertainty intervals resulting from a total of about $60$ state detections per data point. (c) Motional sideband spectra of single atoms after cavity-cooling with $\overline{m}_z = 4.5\pm 2.4$ and (d) after Raman sideband cooling on the (cavity) $z$-axis cooling sideband with $\overline{m}_z = 0.10\pm 0.01$ strongly reducing the sidebands corresponding to $\hat{b}_z$ transitions. The significant presence of the $\hat{b}_z^{\dagger}\hat{b}_y$ sideband shows that the cooling into the motional ground state along the $z$-axis does not effectively cool the motion along the $y$-axis.}
\end{figure}
The results of our measurement, shown in Fig.~\ref{fig:fig2}b, clearly reveal a strong suppression of the carrier transition in the two-photon spectrum. Within the signal-to-noise limits of our data we do not find a discernible carrier contribution, in agreement with the theoretically estimated suppression of the carrier Rabi frequency by a factor $>10^{4}$ limited by gravitation sag (see Appendix \hyperref[AppendixB]{B}). A fit of six Lorentzian curves to the motional sidebands spectrum provides an estimate for the trapping frequencies along the $y$- and $z$-axis. By comparing the relative heights of the heating and cooling motional sidebands we furthermore extract the temperatures of the atoms~\cite{leibfried2003} (see Appendix \hyperref[AppendixC]{C}). The spectral data shown in Fig.~\ref{fig:fig2}c for cavity-cooled atoms yield a mean excitation number of $\overline{m}_z= 4.5\pm 2.4$ quanta along the $z$ direction.   

Next, we implement one-dimensional Raman cooling along the cavity direction. During a $20\,{\rm ms}$ long cooling interval we simultaneously drive the resolved $\hat{b}_z$ cooling sideband (see Fig.~\ref{fig:fig2}a) and apply the optical pumping and repumping light (see Fig. 1c and d). The cooling stage is followed by the recording of a Raman spectrum for temperature determination (Fig.~\ref{fig:fig2}d). We extract steady-state motional excitations $\overline{m}_y= 3.2\pm 0.2$ and $\overline{m}_z= 0.10\pm 0.01$ after Raman cooling. 
\begin{figure} 
\centering
    \includegraphics[width=0.55\columnwidth]{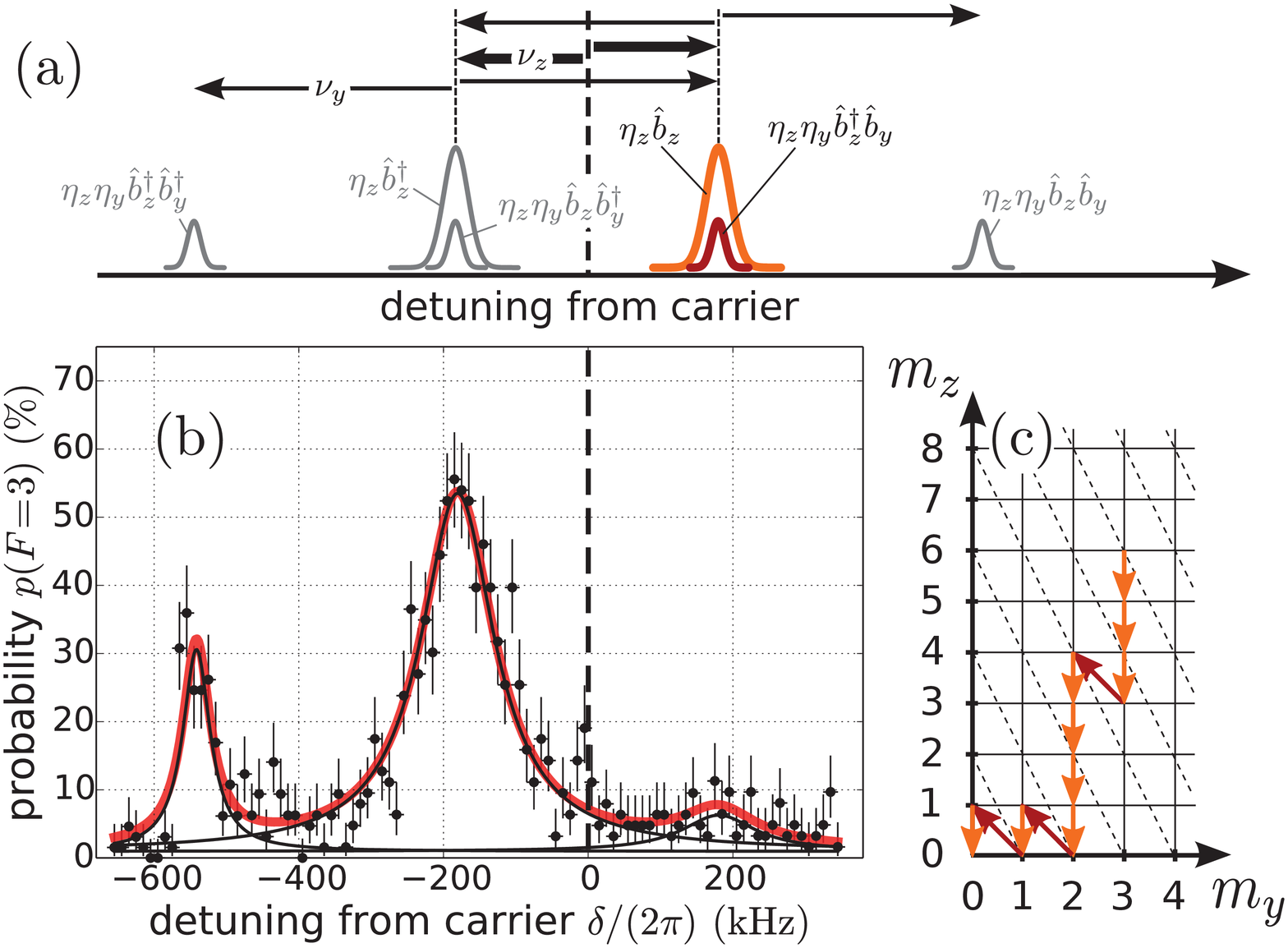}
	\caption[Level Scheme]{\label{fig:fig3}(Color) Continuous two-dimensional Raman cooling. (a) By adjusting the laser powers in the red and the blue dipole trap, the atomic trap frequencies along the $y$- and $z$-axis are matched such that $\nu_y = 2\nu_z$. By addressing the degenerate sidebands atoms are continuously cooled into the two-dimensional motional ground state. (b) The reduced heights of the degenerate $\hat{b}_z$ and $\hat{b}_z^{\dagger}\hat{b}_y$ sidebands in the Raman spectrum demonstrate successful ground state cooling along both dimensions. (c) Schematic of an exemplary cooling trajectory. Dashed lines link points of equal motional energy of the atom.}
\end{figure}

In order to characterize how the cooling process of atoms is influenced by the suppression of the Raman carrier transition, we estimate its fundamental cooling limit~\cite{wineland1979,leibfried2003}. For simplicity we only consider one spatial dimension in the following analysis. During a resolved sideband cooling cycle the atom is driven on the cooling sideband ($\delta= 2\pi \cdot\nu$) from the state ${\ket{\!\uparrow, m }}$ to ${\ket{\!\downarrow, m' = m-1}}$. Resonant single photon excitation by the repumping light and spontaneous decay predominantly return the atom to the state ${\ket{\!\uparrow,m-1}}$ in the Lamb-Dicke regime (see Fig~1c). This pumping of the atom into the ``dark'' motional ground state ${\ket{\!\uparrow,0}}$ is counteracted by heating due to off-resonant Raman excitation in the Lorentzian wing of the carrier and the blue sideband transition. For large ground state occupations the cooling dynamics can be restricted to the ground and the first excited motional state and can be described by the rate equations~\cite{leibfried2003}
\begin{equation}	
\dot{p}_0=p_1 \frac{(\eta \Omega_0)^2}{\Gamma_{\textrm{rep}}}-p_0\left[\cancel{\left(\frac{\Omega_0}{4\pi\nu}\right)^2 \eta^2 \Gamma_{\textrm{rep}}} + \left(\frac{\eta\Omega_0}{8\pi\nu}\right)^2\Gamma_{\textrm{rep}}\right]
\label{eq:eq3}
\end{equation}	
and $\dot{p}_1=-\dot{p}_0$, for the probabilities $p_0$, $p_1$ of the atom to be in the ground and the first excited state. The first term on the right hand side of Eqn.~\ref{eq:eq3} states the rate of the cooling cycle. Resonant Raman excitation on the red sideband is followed by repumping and decay on the carrier at a rate of $\Gamma_{\textrm{rep}}$. The second term, which contains the heating due to off-resonant excitation on the carrier followed by decay on the heating sideband, vanishes in our case due to carrier suppression. The leading heating mechanism is therefore given by off-resonant excitation on the heating sideband and repumping and decay on the carrier described in the third term. This results in a fundamental steady-state mean occupation number $\overline{m}\approx p_1 \approx (\frac{\Gamma_{\textrm{rep}}}{4\pi\nu})^2[\cancel{\,1\,}+1/4]$, which is a factor of 5 smaller than for conventional Raman cooling. However, we do not expect our measurements to explore this fundamental limit, due to the effects of technical heating during the cooling and spectroscopy sequence, as well as due to imperfections in the optical pumping. In addition to improving the cooling limit the carrier suppression should also somewhat relax the initial starting temperature requirements for the onset of robust Raman cooling~\cite{kaufman2012}. Due to the significant contribution from higher order motional sidebands at the border of the Lamb-Dicke regime this effect of carrier suppression will be very small.

In order to extend the Raman cooling from the cavity $z$-axis to the dipole trap $y$-axis one of the motional sidebands ($\hat{b}_z^{\dagger}\hat{b}_y$ or $\hat{b}_z\hat{b}_y$), which couple both directions, needs to be addressed. Simultaneous cooling to the two-dimensional motional ground state is achieved in the experiment by adjusting the trapping frequencies to satisfy $\nu_y=2\nu_z$. The two degenerate  $\hat{b}_z^{\dagger}\hat{b}_y$ and $\hat{b}_z$ sidebands are addressed with a $20\,$ms long Raman, repump and optical pumping pulse and continuously cooled into the two-dimensional ground state (see Fig~3). From the height of the Raman sidebands after cooling we estimate upper limits for the mean excitation numbers of $\overline{m}_y^{(\rm max)}= 0.3\pm 0.2$ and $\overline{m}_z^{(\rm max)}= 0.11\pm 0.05$ respectively. The two-dimensional ground state cooling can further be extended to all three dimensions if sufficiently strong confinement and momentum transfer are also achieved along the $x$-axis. Many different experimental scenarios can be envisioned for this purpose, such as another phase-locked blue detuned standing wave along the $x$-direction, or an additional red detuned standing wave and an additional Raman beam along the $x$-direction~\cite{Note3}.           
\begin{figure}
\centering
  \includegraphics[width=0.5\columnwidth]{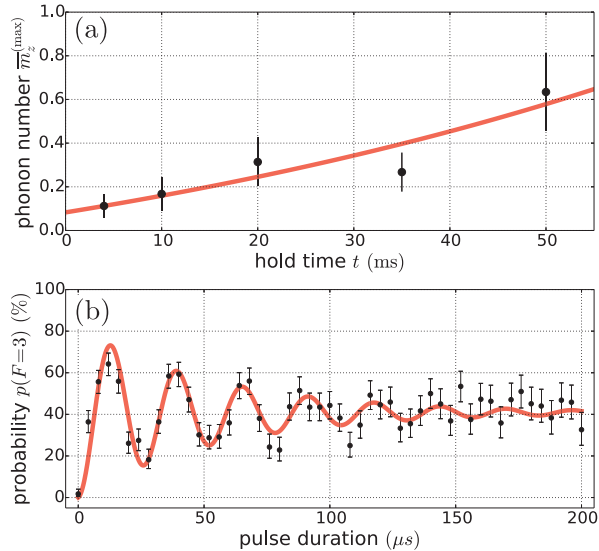} 
\caption[RabiOsci]{\label{fig:fig4}(Color) (a) Measurement of the heating rate along the $z$-axis (cavity) direction. The mean motional quantum number is recorded as a function of the hold time in the optical potential. The solid line fits the data with an exponential heating model that considers parametric heating due to fluctuations in the intra-cavity dipole trap power (see text for details). (b) Coherent Rabi-oscillations on the $z$-axis motional sideband.}
\end{figure}

The heating and position jumping of atoms in optical lattices formed by optical cavities have been noted in the past~\cite{maunz2004,khudaverdyan2008}. Compared to standing wave potentials generated by counter-propagating running wave laser beams the resonantly enhanced dipole field inside Fabry-P\'{e}rot cavities is additionally affected by the noise of the cavity lock. Considering trap intensity fluctuations as the primary contribution, parametric heating in the approximately harmonic trap should lead to a linear increase of the heating rate with motional energy~\cite{gehm1998}. We directly measure the mean motional quantum number of single atoms $\overline{m}_z ^{({\rm max})}(t)$ as a function of the hold time in the optical potential (see Fig.~\ref{fig:fig4}a). To account for the expected exponential increase in motional energy we fit the data with $\overline{m}_z ^{({\rm max})}(t) = [\overline{m}_z ^{({\rm max})}(t=0)+1/2]\cdot \exp(\Gamma_H t)-1/2$, which results in an initial maximum phonon number along the $z$-direction of $\overline{m}_z ^{({\rm max})}(t=0) = 0.08 \pm 0.06$ and a heating rate constant of $\Gamma_H = (12\pm 3)\,{\rm Hz}$. For these cold starting temperatures the low heating rate of less than one motional quantum in $50\,$ms allows atoms to remain well-localized during the timescales of most quantum optics experiments. Indeed, our experiment is in fact limited by spin relaxation due to spontaneous Raman scattering processes at the $100\,$ms timescale. 

Finally we study coherent dynamics on the motional sideband. For this purpose an initial Raman cooling interval prepares the atom with high probability in the state ${\ket{\!\uparrow,m_z = 0, m_y=0}}$. The excitation probability as a function of Raman pulse length on the $\hat{b}^{\dagger}_z$ heating sideband shows Rabi oscillation with a Rabi frequency of $2\pi \cdot (38.2\pm0.4)\:$kHz and a decay constant of $(55\pm 7)\,\mu$s (see Fig.~\ref{fig:fig4}). Earlier measurements of carrier Rabi oscillations under similar experimental conditions but driven by microwave pulses have shown coherence times in excess of $100\,\mu$s. We attribute the reduced coherence time to effects associated with the oscillation of the atom along the weakly confined $x$-direction. Since $\nu_z$ directly depends on the position along the $x$-direction, this motion strongly influences the sideband transitions.

\section{Conclusion}

The measurements presented here implement and characterize a method that should be valuable to a range of atomic physics experiments. It provides a robust experimental solution for ever more integrated and miniaturized setups, which make the fast and lossless preparation of cold atoms a significant challenge. We highlight that the absence of the carrier is a generic feature of any scheme that traps atoms in the zero-crossing of the electric field of one of the two Raman beams. Carrier-free Raman manipulation is therefore suitable for many blue detuned optical dipole potentials, including optical lattices, microtrap arrays and higher order paraxial (e.g. ``doughnut'') beams~\cite{kuga1997}. In the context of cavity coupled atoms Raman cooling has the advantage of being readily applicable to more than one particle, in contrast to most cavity cooling schemes. Compared to previous cooling experiments with additional Raman beams along the cavity direction, the presented scheme provides constant Raman coupling conditions for atoms at different axial positions inside the cavity.

\addcontentsline{toc}{section}{Acknowledgments}

We thank T. Kampschulte for discussions and for contributions to the early stage of the experiment. The work has been supported by the Bundesministerium f\"ur Forschung und Technologie (BMFT, Verbund Q.com-Q), and by funds of the European Commission training network CCQED and the integrated project SIQS. 

\renewcommand\thefigure{A\arabic{figure}}   
\renewcommand\theequation{A\arabic{equation}}    
\setcounter{equation}{0} 
\setcounter{figure}{0} 
\section*{Appendix A. State detection and cavity cooling}
\label{AppendixA}
Fast and non-destructive readout of the hyperfine state of single atoms is performed in the experiment by cavity assisted detection~\cite{boozer2006, bochmann2010}. The high-finesse cavity is stabilized to the lock laser light such that one of the cavity modes is $20\,$MHz blue detuned from the $\ket{F=4}\rightarrow\ket{F'=5}$ transition of the atomic D$_2$-line. A weak probe laser beam, resonant with this mode, is transmitted through the cavity and its intensity is monitored by a single photon counter module. The lock and probe light are separated by three free spectral ranges in order that an atom near the center of the cavity trapped in the node of the blue detuned lock light is situated at the anti-node of the probe field. 
The vacuum Rabi splitting caused by the strong photon-atom coupling for a single atom in the $F=4$ manifold leads to a suppression of the transmitted photon count rate, whereas for an atom in the $F=3$ manifold the count rate remains unchanged relative to the empty cavity. Typical detection fidelities $>95\,\%$ are reached within $2\,{\rm ms}$.  
The illumination of atoms with probe light that is $20\,{\rm MHz}$ blue-detuned relative to the atomic resonance furthermore gives rise to cavity cooling effects~\cite{bienert2012} that keep the atoms at a mean motional excitation $\overline{m}_z = 4.5\pm 2.4$ along the cavity axis.
\renewcommand\thefigure{B\arabic{figure}}   
\renewcommand\theequation{B\arabic{equation}}    
\setcounter{equation}{0} 
\setcounter{figure}{0} 
\section*{Appendix B. Limits to carrier suppression}
\label{AppendixB}
The suppression of the carrier transition is caused by the zero-crossing of the electric field amplitude of one of the Raman lasers at the equilibrium position of the atom. Since this blue detuned Raman laser light is also the source of atomic confinement, atoms will localize at the minimum of the light intensity. However, offset Raman light intensity at the trap minimum, for example, due to the imperfect intensity balance of the two interfering beams forming an optical standing wave, will give rise to a residual carrier Raman coupling. In addition, external forces displacing the trap center from the intensity minimum can lead to a further increase of the carrier contribution. 

In order to quantify these effects for our system we define the suppression factor 
\begin{equation}
\mathcal{S}= \frac{\Omega_{\uparrow m_y\:m_z=0,\downarrow m_y\:m_z'=1}}{\Omega_{\uparrow m_ym_z,\downarrow m_ym_z }}\frac{1}{\eta_z}
\label{eq:eqA1}
\end{equation} as the ratio of the ground state blue-sideband Rabi frequency to the carrier Rabi frequency along the $z$-axis, which we normalize by the Lamb-Dicke factor. Thus, in the Lamb-Dicke regime Eqn.~\ref*{eq:eqA1} yields $\mathcal{S} = 1$ for an atom addressed by two independent running wave Raman lasers and $\mathcal{S} \rightarrow \infty$ for perfect carrier suppression. 

If we assume that the Raman standing wave is formed by counter-propagating beams with an intensity ratio $\mathcal{R}\leq 1$, we obtain at the position of the atom the expression
\begin{align}
\label{eq:eqA2}
e^{ik_z \hat{z}}-\sqrt{\mathcal{R}}&\,e^{-ik_z \hat{z}}  \\
&=(1-\sqrt{\mathcal{R}})\,e^{ik_z \hat{z}}+2i\sqrt{\mathcal{R}}\,\sin(k_z \hat{z})\notag \\ 
&\approx (1-\sqrt{\mathcal{R}})\hat{\mathds{1}}_{z}  + i(1+\sqrt{\mathcal{R}}) \,\eta_z (\hat{b}_z^{\dagger}+\hat{b}_z)\notag
\end{align}
for the motional coupling along the $z$-axis in analogy to Eqn.~\ref{eq:eq2}. The suppression factor is then given by 
$\mathcal{S}=\frac{1+\sqrt{\mathcal{R}}}{1-\sqrt{\mathcal{R}}}$. Our Fabry-P\'{e}rot cavity with high reflectivity mirrors ($\mathcal{R}\approx 1$) has a finesse of $\mathcal{F}=\frac{\pi\sqrt{\mathcal{R}}}{1-\mathcal{R}}\approx 10^6$ and we therefore expect the carrier suppression to be limited at the $\mathcal{S} \approx \frac{4}{1-\mathcal{R}}\approx \frac{4\mathcal{F}}{\pi} \approx 10^6$ level due to the ``running wave'' component of the cavity mode. 

A more stringent limit, however, is caused by gravitational acceleration $g$ in the vertically orientated optical lattice with trap frequency $\nu_z$. At a vertical position shift of $z_s=g/(2\pi\nu_z)^2 \approx 6\,$pm the atom experiences a field proportional to $\sin(k_z z)$, which can be Taylor expanded around the point $z=z_s$ and leads to a motional state coupling of 
\begin{align}
\label{eq:eqA3}
\sin(k_z \hat{z}) &\approx \sin(k_z z_s)\hat{\mathds{1}}_{\tilde{z}} +\cos(k_z z_s)\;k_z \hat{\tilde{z}}\\
&= \sin(k_z z_s)\hat{\mathds{1}}_{\tilde{z}} +\cos(k_z z_s) \,\eta_z (\hat{b}_{\tilde{z}}^{\dagger}+\hat{b}_{\tilde{z}}),\notag
\end{align}
where $\tilde{z}=z-z_s$.
The gravitational sag therefore limits the suppression to $\mathcal{S} = \frac{\cos(k_z z_s)}{\sin(k_z z_s)} \approx 1/(k_z z_s) \approx 2\times 10^4$.    
\renewcommand\thefigure{C\arabic{figure}}   
\renewcommand\theequation{C\arabic{equation}}    
\setcounter{equation}{0} 
\setcounter{figure}{0} 
\section*{Appendix C. Temperature estimation in two-dimensions}
\label{AppendixC}
\begin{figure}[b]
\centering
  \includegraphics[width=0.5\columnwidth]{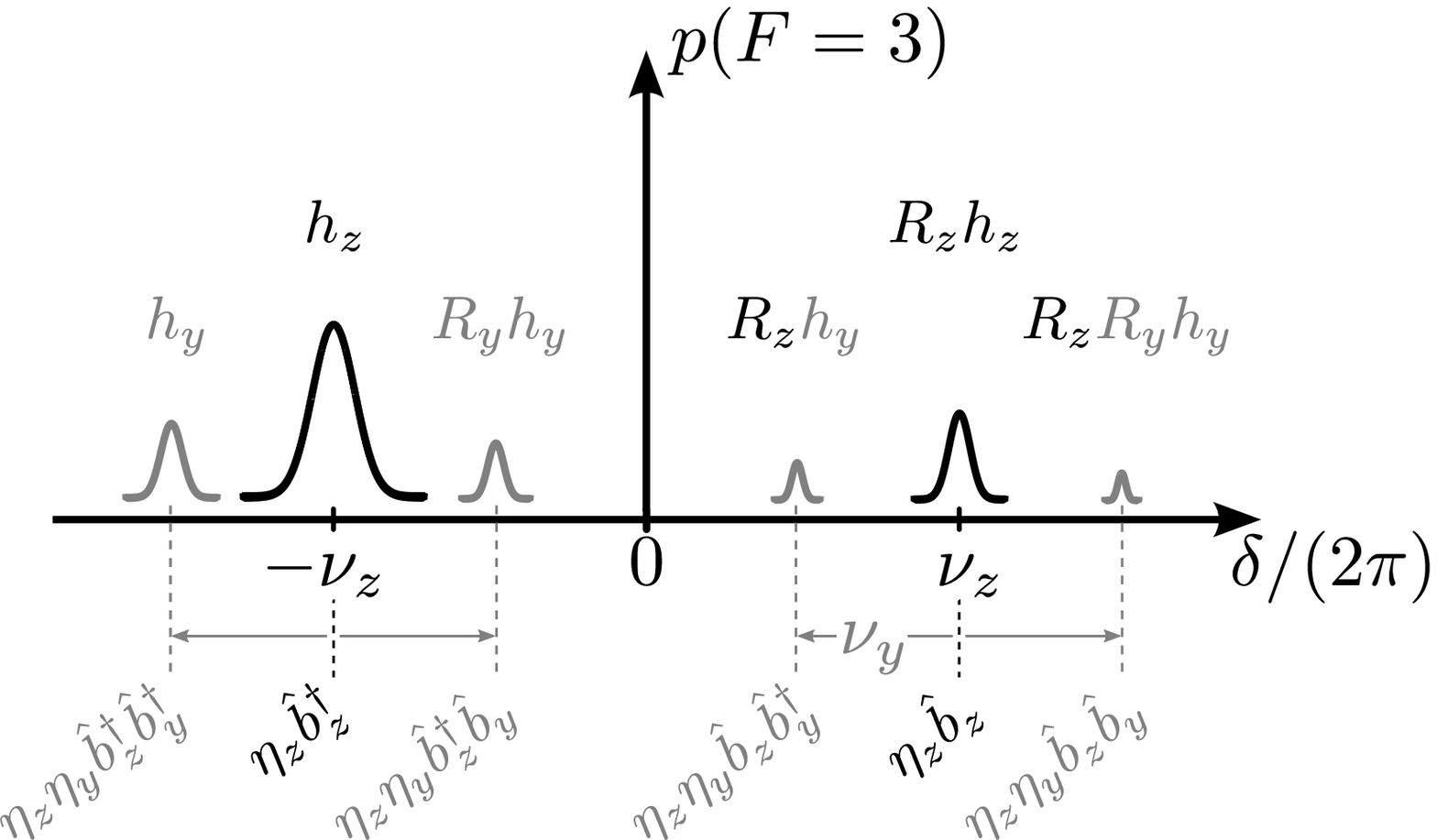} 
\caption[Spectrum2DTemperature]{\label{fig:figC1} The relative heights of motional sidebands in the two-photon spectrum as a function of the atomic temperatures along both trap axes. See text for details.}
\end{figure}
The relative height of the heating and cooling motional sidebands remains a robust measure of atomic temperatures also for sidebands coupling motion in two-dimensions.   
Two independent Boltzmann-distributions are assumed for the level populations of motional states $m_i$ along the $y$- and $z$-axis
\begin{equation}
p_{m_i}= \frac{1}{\overline{m}_i+1}\left(\frac{\overline{m}_i}{\overline{m}_i+1}\right)^{m_i}, \quad\quad i=y,z.
\label{eq:eqA4}
\end {equation} 
The transfer probability of an atom initially prepared in ${\ket{\!\uparrow}}$ to the opposite spin state ${\ket{\!\downarrow}}$ after applying a Raman pulse of duration $t$ is given by
\begin{equation}
p(t) =\sum\limits_{m_y=0}^{\infty}\sum\limits_{m_z=0}^{\infty} p_{m_y} p_{m_z} \sin^2(\Omega_{\uparrow m_y m_z,\downarrow m_y' m_z'}\;t/2).
\label{eq:eqA5}
\end{equation}  
We determine the ratio $R_z$ of the transition probabilities on the $z$-axis cooling ($\hat{b}_z$) and heating ($\hat{b}_z^{\dagger}$) motional sideband according to~\cite{leibfried2003}
\begin{align}
R_z&=\frac{\sum\limits_{m_y=0}^{\infty} \sum\limits_{m_z=1}^{\infty} p_{m_y} p_{m_z} \sin^2(\Omega_{\uparrow m_y m_z,\downarrow m_y m_z-1}\;t/2)}{\sum\limits_{m_y=0}^{\infty} \sum\limits_{m_z=0}^{\infty} p_{m_y} p_{m_z} \sin^2(\Omega_{\uparrow m_y m_z,\downarrow m_y m_z+1}\;t/2)}\notag \\ 
&=\frac{\sum\limits_{m_y=0}^{\infty} \sum\limits_{m_z=0}^{\infty} p_{m_y} p_{m_z+1} \sin^2(\Omega_{\uparrow m_y m_z+1,\downarrow m_y m_z}\;t/2)}{\sum\limits_{m_y=0}^{\infty} \sum\limits_{m_z=0}^{\infty} p_{m_y} p_{m_z} \sin^2(\Omega_{\uparrow m_y m_z,\downarrow m_y m_z+1}\;t/2)} \notag\\ 
&=\frac{\overline{m}_z}{\overline{m}_z+1},
\label{eq:eqA6}
\end{align}
where we have used $\Omega_{\uparrow m_y m_z,\downarrow m_y' m_z'}=\Omega_{\uparrow m_y' m_z',\downarrow m_y m_z}$ and $p_{{m_i}+1} = \frac{\overline{m}_i}{\overline{m}_i+1}p_{{m_i}}$ in the last step. 

Analogous calculations yield the ratio of transitions involving the decrease and transitions involving the increase of one motional quantum along the $y$-axis 
\begin{equation}
R_y =\frac{\overline{m}_y}{\overline{m}_y+1},
\label{eq:eqA7}
\end{equation}    
and they result in the relative scaling of all sidebands summarized in Fig.~\ref{fig:figC1}. The height factors $h_y$, $h_z$ depend on the experimental details of the chosen Rabi spectroscopy pulse and do not play a role in the determination of the mean excitation numbers 
\begin{equation}
\overline{m}_y = \frac{R_y}{1-R_y} \quad \textrm{and}  \quad \overline{m}_z = \frac{R_z}{1-R_z}.
\label{eq:eqA8}
\end{equation} 
This analysis rigorously only applies to resolved sidebands. For the temperature estimation with degenerate sidebands we estimate an upper limit for the individual mean excitation numbers $\overline{m}_y^{(\textrm{max})}$ and $\overline{m}_z^{(\textrm{max})}$ by attributing the height of the cooling sideband in its entirety to each motional axis. 
\end{document}